\documentstyle[12pt,epic,eepic,epsfig,rotate]{article}
\title{Microscopic Foundation of Stochastic Game Dynamical Equations}
\author{Dirk Helbing}
\begin{document}
\maketitle
\section*{Abstract}
The game dynamical equations are derived from 
Boltzmann-like equations for individual pair interactions by assuming
a certain kind of imitation behavior, the so-called proportional 
imitation rule. They can be extended to a stochastic 
formulation of evolutionary game theory which allows the derivation
of approximate and corrected mean value and covariance equations.
It is shown that, in the case of phase transitions (i.e. multi-modal 
probability distributions), the mean value equations do not agree with the 
game dynamical equations. Therefore, their exact meaning is carefully
discussed. Finally, some generalizations of the behavioral model are presented,
including effects of expectations,
other kinds of interactions, several subpopulations, or memory effects.

\section{Introduction} \label{intro}

Since von Neumann and Morgenstern have initiated
the field of game theory,\footnote{J. von Neumann and O. Morgenstern,
{\it Theory of Games and Economic Behavior}. Princeton: Princeton University
Press 1944.} it has often proved of great value for the quantitative 
description and understanding of the competition and co-operation between 
individuals. Game theory focusses on two questions:
1. Which is the optimal strategy in a given situation?
2. What is the dynamics of strategy choices in cases of repeatedly interacting
individuals? In this connection game dynamical equations\footnote{P. Taylor
and L. Jonker, ``Evolutionarily Stable Strategies and Game Dynamics'', in:
{\it Math. Biosciences} 40, 1978, pp. 145--156.\protect\\ 
J. Hofbauer and K. Sigmund, {\it Evolutionstheorie und dynamische Systeme}. Berlin: Parey 1984.} 
find a steadily increasing interest.
Although they agree with the replicator equations
of evolution theory (cf. Sec. 2), they cannot be substantiated 
in the same way. Therefore,
we will be looking for a foundation of the game dynamical equations which bases
on individual actions and decisions (cf. Sec. 4). 
\par
In addition, we will formulate a stochastic version of evolutionary
game theory (cf. Sec. 3). This allows to investigate the effects of fluctuations on the
dynamics of social systems. In order to illustrate the essential ideas,
a concrete model for the self-organization of
behavioral conventions is presented (cf. Sec. 5).
We will see that the game dynamical equations
describe the average evolution of social systems only for a certain
time period. Therefore, a criterium for their validity will be 
developed (cf. Sec. 6). 
Finally, we will present possible extensions to more general behavioral models
and discuss the actual meaning of the game dynamical equations (cf. Sec. 7).

\section{The Game Dynamical Equations} 

Let $p_x(t)$ with
\begin{equation}
 0 \le p_x(t) \le 1 \qquad \mbox{and} \qquad \sum_x p_x(t) = 1
\end{equation}
denote the {\it proportion} of individuals pursuing the 
{\it behavioral strategy} $x \in S$ at time $t$. We assume the considered 
strategies to be mutually exclusive. The set $S$ of strategies may be
discrete or continuous, finite or infinite. The only difference will be
that sums over $x$ are to be replaced by integrals in cases of 
continuous sets. By $A_{xy}$ we will denote the possibly time-dependent
{\it payoff} of an individual using strategy $x$ 
when confronted with an individual
pursuing strategy $y$. Hence, his/her {\it expected success} 
$\langle E_x \rangle_t$ 
will be given by the weighted mean value
\begin{equation}
 \langle E_x \rangle_t =  \sum_y A_{xy} \, p_y(t) \, ,
\end{equation}
since $p_y$ is the probability that the interaction partner uses strategy $y$.
In addition, the {\it average expected success} will be
\begin{equation}
 \overline{\langle E \rangle_t} = \sum_x p_x(t) \langle E_x \rangle_t 
 = \sum_x \sum_y p_x(t) A_{xy} \, p_y(t) \, .
\end{equation}
\par
Assuming that the relative temporal increase $(dp_x/dt)/p_x$
of the proportion $p_x$ of individuals pursuing strategy $x$ is proportional
to the difference between the expected success $\langle E_x \rangle_t$ and
the average expected success $\overline{\langle E \rangle_t}$, we obtain the
{\it game dynamical equations}
\begin{eqnarray}
 \frac{dp_x(t)}{dt} &=& \nu p_x(t) \big[ \langle E_x \rangle_t - \overline{\langle E \rangle_t} \big] 
 \nonumber \\
 &=& \nu p_x(t) \Big[ \langle E_x \rangle_t 
- \sum_y p_y(t) \langle E_y \rangle_t \Big] \, ,
\label{repli}
\end{eqnarray}
where the possibly time-dependent proportionality factor $\nu$ is a measure for
the {\it interaction rate} with other individuals. 
According to (\ref{repli}), the proportions of strategies with an above-average
success $\langle E_x \rangle_t > \overline{\langle E \rangle_t}$ increase, whereas the other strategies will be 
diminished. Note, that the proportion of a strategy does not
necessarily increase or decrease monotonically. Certain payoffs
are related with an {\it oscillatory} or even {\it chaotic} 
dynamics\footnote{W. Schnabl, P. F. Stadler, C. Forst, and P. Schuster,
``Full Characterization of a Strage Attractor. Chaotic Dynamics in
Low-Dimensional Replicator Systems'', in: {\it Physica D} 48, 1991,
pp. 65--90.}.
\par
Equations (\ref{repli}) are identical with the replicator equations
from evolutionary biology. They can be extended to the {\it selection-mutation
equations}
\begin{eqnarray}
 \frac{dp_x(t)}{dt} &=& \nu p_x(t) \Big[ \langle E_x \rangle_t 
 - \sum_y p_y(t) \langle E_y \rangle_t \Big] \nonumber \\
 &+& \sum_y \big[ p_y(t) w_1(y\rightarrow x) 
 - p_x(t) w_1(x\rightarrow y) \big] \, .
\label{mutation}
\end{eqnarray}
The terms which agree with (\ref{repli}) describe a selection of
superior strategies. The new terms correspond to the effect 
of mutations, i.e. to {\it spontaneous} changes from strategy $x$ to other
strategies $y$ with possibly time-dependent {\it transition rates} 
$w_1(x \rightarrow y)$ (last term) and the inverse transitions. They allow to
describe {\it trial and error behavior} or behavioral fluctuations.
  
\section{Stochastic Dynamics: The Master Equation}

Let us consider a social system consisting of a constant number
\begin{equation}
 N = \sum_x n_x(t)
\label{sum}
\end{equation}
of individuals. Herein, $n_x(t)$ denotes the number of individuals
who pursue strategy $x$ at time $t$. Hence, the time-dependent vector
\begin{equation}
 \vec{n} = (n_1,n_2, \dots, n_x, \dots, n_y, \dots )
\end{equation}
reflects the {\it strategy distribution} in the social system
and is called the {\it socioconfiguration}. If the individual strategy changes
are subject to random fluctuations (e.g. due to trial and error
behavior or decisions under uncertainty), 
we will have a stochastic dynamics. Therefore, 
given a certain socioconfiguration $\vec{n}_0$ at time $t_0$, for the 
occurence of the strategy distribution $\vec{n}$ at a time $t > t_0$ 
we can only calculate a certain
{\it probability} $P(\vec{n},t)$. Its temporal change $dP/dt$ is
governed by the so-called {\it master equation}\footnote{W. Weidlich and
G. Haag, {\it Concepts and Models of a Quantitative Sociology. The Dynamics of
  Interacting Populations.} Berlin: Springer, 1983.}
\begin{equation}
 \frac{dP(\vec{n},t)}{dt} = \sum_{\vec{n}^{\,\prime}} \big[
 P(\vec{n}',t) W(\vec{n}^{\,\prime} \rightarrow \vec{n})
 -  P(\vec{n},t) W(\vec{n} \rightarrow
 \vec{n}^{\,\prime}) \big] \, .
\label{master}
\end{equation} 
The sum over $\vec{n}^{\,\prime}$ extends over all socioconfigurations
fulfilling $n_x \in \{0,1,2,\dots\}$ and (\ref{sum}).
\par
According to equation (\ref{master}), an {\it increase} of
the probability $P(\vec{n},t)$ of having socioconfiguration $\vec{n}$
is caused by transitions from other socioconfigurations
$\vec{n}^{\,\prime}$ to $\vec{n}$. While a {\it decrease} of $P(\vec{n},t)$ 
is related to changes from $\vec{n}$ to other 
socioconfigurations $\vec{n}^{\,\prime}$. The corresponding
changing rates are proportional to the {\it configurational transition rates} 
$W(\vec{n}\rightarrow \vec{n}^{\,\prime})$ of changes to socioconfigurations
$\vec{n}^{\,\prime}$ {\it given} the socioconfiguration $\vec{n}$ and
to the probability $P(\vec{n},t)$
of {\it having} socioconfiguration $\vec{n}$ at time $t$.
\par
The configurational transition rates $W$ have the meaning of transition probabilities per
time unit and must be non-negative quantities. Frequently, 
the individuals can be assumed to change their strategies independently 
of each other. 
Then, the configurational transition rates have the form
\begin{equation}
 W(\vec{n}\rightarrow \vec{n}^{\,\prime})
 = \left\{ \begin{array}{ll}
 n_x w(x \rightarrow y; \vec{n}) & \mbox{if } \vec{n}^{\,\prime} = 
 \vec{n}_{xy} \\
 0 & \mbox{otherwise,}
\end{array} \right.
\label{trans}
\end{equation}
i.e. they are proportional to the number $n_x$ of individuals who
may change their strategy from $x$ to another strategy $y$ with an
{\it individual transition rate} $w(x \rightarrow y;\vec{n}) \ge 0$.
In relation (\ref{trans}), the abbreviation
\begin{equation}
 \vec{n}_{xy} = (n_1, n_2, \dots, n_x - 1, \dots ,n_y + 1, \dots ) 
\end{equation}
means the socioconfiguration
which results after an individual has changed his/her strategy from
$x$ to $y$.
\par 
It can be shown that the master equation has the properties 
\begin{equation}
 P(\vec{n},t) \ge 0 \qquad \mbox{and} \qquad
 \sum_{\vec{n}} P(\vec{n},t) = 1 
\end{equation}
for all times $t$, if they are fulfilled at some initial time $t_0$.
Therefore, the master equation actually describes the temporal evolution
of a probability distribution. 

\section{Approximate Mean Value Equations} 

In order to connect the stochastic model to the game dynamical equations,
we must specify the individual transition rates $w$ in a suitable way. Therefore, we
derive the mean value equations related to the master equation (\ref{master})
and compare them to the selection-mutation equations (\ref{mutation}).
\par
The proportion $p_x$ is defined as the {\it mean value}
\begin{equation}
 \langle f \rangle_t = \sum_{\vec{n}} f(\vec{n},t) P(\vec{n},t) 
\label{mv}
\end{equation}
of the number $f(\vec{n},t) = n_x$ 
of individuals pursuing strategy $x$, divided by the total number $N$ of
considered individuals:
\begin{equation}
 p_x(t) = \frac{\langle n_x \rangle_t}{N}
 = \frac{1}{N} \sum_{\vec{n}} n_x P(\vec{n},t) \, .
\label{av}
\end{equation}
Taking the time derivative of $\langle n_x \rangle_t$ and inserting the master
equation gives
\begin{eqnarray}
 \frac{d\langle n_x \rangle_t}{dt} &=&
 \sum_{\vec{n}} n_x \big[
 P(\vec{n}',t) W(\vec{n}^{\,\prime} \rightarrow \vec{n})
 -  P(\vec{n},t) W(\vec{n} \rightarrow \vec{n}^{\,\prime}) \big] \nonumber \\
 &=&  \sum_{\vec{n}} (n'_x - n_x) 
 W(\vec{n} \rightarrow \vec{n}^{\,\prime}) P(\vec{n},t)\, ,
\label{not}
\end{eqnarray}
where we have interchanged $\vec{n}$ and $\vec{n}^{\,\prime}$ in the
first term on the right hand side. Taking into account relation (\ref{trans}),
we get
\begin{eqnarray}
 \frac{d\langle n_x \rangle_t}{dt} &=&
 \sum_{\vec{n}_{yx}} 
 n_y w(y\rightarrow x;\vec{n}) P(\vec{n},t)
 - \sum_{\vec{n}_{xy}}  n_x w(x\rightarrow y;\vec{n}) P(\vec{n},t) \nonumber \\
 &=&  \sum_y \big[ n_y w(y\rightarrow x;\vec{n}) - n_x w(x\rightarrow y;\vec{n})
 \big] P(\vec{n},t) \, .
\end{eqnarray}
With (\ref{av}) this finally leads to the {\it approximate mean value
equations}
\begin{equation}
 \frac{dp_x(t)}{dt} = \sum_y \big[ 
 p_y(t) w(y\rightarrow x;\langle \vec{n} \rangle_t) 
 - p_x(t) w(x\rightarrow y;\langle \vec{n} \rangle_t) \big] 
\label{rate}
\end{equation}
However, these are only exact, if the individual
transition rates $w$ are independent of the
socioconfiguration $\vec{n}$. Anyhow, they are {\it approximately} 
valid as long as the probability distribution $P(\vec{n},t)$ is narrow, so that
the mean value $\langle f(\vec{n},t) \rangle_t$ of a function $f(\vec{n},t)$
can be replaced by the function $f(\langle \vec{n} \rangle_t,t)$ of the
mean value. This problem will be discussed in detail later on.
\par
Comparing the rate equations (\ref{rate}) with
the selection-mutation equations (\ref{mutation}), we find a complete
correspondence for the case
\begin{equation}
 w(y \rightarrow x;\vec{n}) = w_1(y \rightarrow x) + 
 w_2(y \rightarrow x) \, n_x
\label{add}
\end{equation}
with
\begin{equation}
 w_2(y \rightarrow x) = \frac{\nu}{N} \max(E_x - E_y,0) 
\label{prop}
\end{equation}
and the {\em success}
\begin{equation}
 E_x = \sum_y A_{xy} \, \frac{n_y}{N} \, ,
\label{es}
\end{equation}
since
\begin{equation}
 \max(\langle E_x \rangle_t - \langle E_y \rangle_t,0) 
 - \max(\langle E_y \rangle_t - \langle E_x \rangle_t,0) 
 = \langle E_x \rangle_t - \langle E_y \rangle_t \, .
\end{equation}
Whereas $w_1$ is again the mutation rate (i.e. the rate of spontaneous
transitions), the additional term in (\ref{add}) describes {\it imitation 
processes,} where individuals take over the strategy $x$ of their respective
interaction partner. Imitation processes correspond to pair interactions 
of the form
\begin{equation}
 y + x \rightarrow x + x \, .
\end{equation}
Their frequency is proportional to the number $n_x$ of interaction partners
who may convince an individual of strategy $x$.
The proportionality factor $w_2$ is the {\it imitation rate}. 
\par
Relation (\ref{prop}) is called the {\it proportional imitation 
rule} and can be shown to be the best
learning rule.\footnote{K. H. Schlag, ``Why Imitate, and if so, How? A Bounded
Rational Approach to Multi-Armed Bandits'', Discussion Paper No. B-361,
Department of Economics, University of Bonn.}
It was discovered in 1992\footnote{D. Helbing, ``Interrelations between
Stochastic Equations for Systems with Pair Interactions'', in {\it Phy\-si\-ca A}
181, 1992, pp. 29--52.}
and says that an imitation behavior only takes place, if the strategy
$x$ of the interaction partner turns out to have a greater 
success $E_x$ than the own strategy $y$. In such cases, the imitation
rate is proportional to the difference 
$(E_x - E_y)$ between the success' of the alternative $x$ and the
previous strategy $y$, i.e. strategy changes
occur more often the greater the advantage of the new strategy $x$ would be.
\par
All specifications of the type
\begin{equation}
 w_2(y \rightarrow x) = C +
 \frac{\nu}{N} \big[ \lambda E_x - (1-\lambda) E_y \big] 
\label{prop2}
\end{equation}
with an arbitrary parameter $\lambda$ also lead to the game
dynamical equations. However, individuals would then, with a certain rate,
take over the strategy $x$ of the interaction partner, even if its
success $E_x$ is smaller than that of the previously used strategy $y$.
Moreover, if $C$ is not chosen sufficiently large, the 
individual transition rates $w\ge 0$ can become negative.
\par
In summary, we have found a microscopic foundation of evolutionary game 
theory which bases on four plausible assumptions: 
1. Individuals evaluate the success of a strategy as its average 
payoff in interactions with other individuals (cf. (\ref{es})).
2. They compare the success of their strategy with that
of the respective interaction partner, basing on observations or an exchange 
of experiences.
3. Individuals imitate each others behavior.
4. In doing so, they apply the proportional imitation rule (\ref{prop}) [or
(\ref{prop2})].

\section{Self-Organization of Behavioral Conventions}

For illustrative reasons, we will now discuss an example which allows
to understand how social conventions emerge. We consider the simple case
of two alternative strategies $x \in \{1,2\}$ and assume them to be
equivalent so that the payoff matrix is symmetrical:
\begin{equation}
 \Big( A_{xy} \Big) = \left(
\begin{array}{cc}
 A+B & B \\
 B & A+B
\end{array} \right) \, .
\end{equation}
If $A>0$, the additional payoff $A$ reflects the {\it advantage} of using 
the same strategy like
the respective interaction partner. This situation is, for example, 
given in cases of network externalities like in the historical rivalry between
the video systems VHS and BETA MAX\footnote{W. B. Arthur, ``Competing
  Technologies, Increasing Returns, and Lock-In by Historical Events'', in
{\it The Economic Journal} 99, 1989, pp. 116--131.}. Finally, the
mutation rates are taken constant, i.e. $w_1(x\rightarrow y) = W_1$.
\par
The resulting game dynamical equations are
\begin{equation}
 \frac{dp_x(t)}{dt} = -2 \left[p_x(t) - \frac{1}{2}\right] 
 \Big\{ W_1 + \nu A p_x(t) 
 \big[ p_x(t) - 1 \big] \Big\} \, .
\label{game}
\end{equation}
Obviously, they have only {\it one} stable stationary solution if the
(control) parameter
\begin{equation}
 \kappa = 1 - \frac{4W_1}{\nu A}
\end{equation}
is smaller than zero. However, for $\kappa > 0$ equation (\ref{game})
can be rewritten in the form
\begin{equation}
 \frac{dp_x(t)}{dt} = -2\nu A \left[p_x(t) - \frac{1}{2} \right]
 \left[ p_x(t) - \frac{1 + \sqrt{\kappa}}{2} \right]
 \left[ p_x(t) - \frac{1 - \sqrt{\kappa}}{2} \right] \, .
\end{equation}
The stationary solution $p_x = 1/2$ is unstable, then, but
we have two new stable stationary solutions $p_x = (1/2 \pm
\sqrt{\kappa}/2)$. That is, dependent on the detailled initial condition,
one strategy will win the majority of users although both strategies are
completely equivalent. This phenomenon is called {\it symmetry breaking}.
It will be suppressed, if the mutation rate $W_1$ is larger than 
the advantage effect $\nu A/4$.
\par
The above model allows to understand how behavioral conventions come about.
Examples are the pedestrians' preference for the right-hand side
(in Europe), the revolution direction of clock hands, the
direction of writing, or the already mentioned triumph of the video 
system VHS over BETA MAX.
\par
\begin{figure}[htbp]
\parbox[b]{7cm}{                                                   
\epsfxsize=7cm 
\centerline{\rotate[r]{\hbox{\epsffile[28 28 570                              
556]{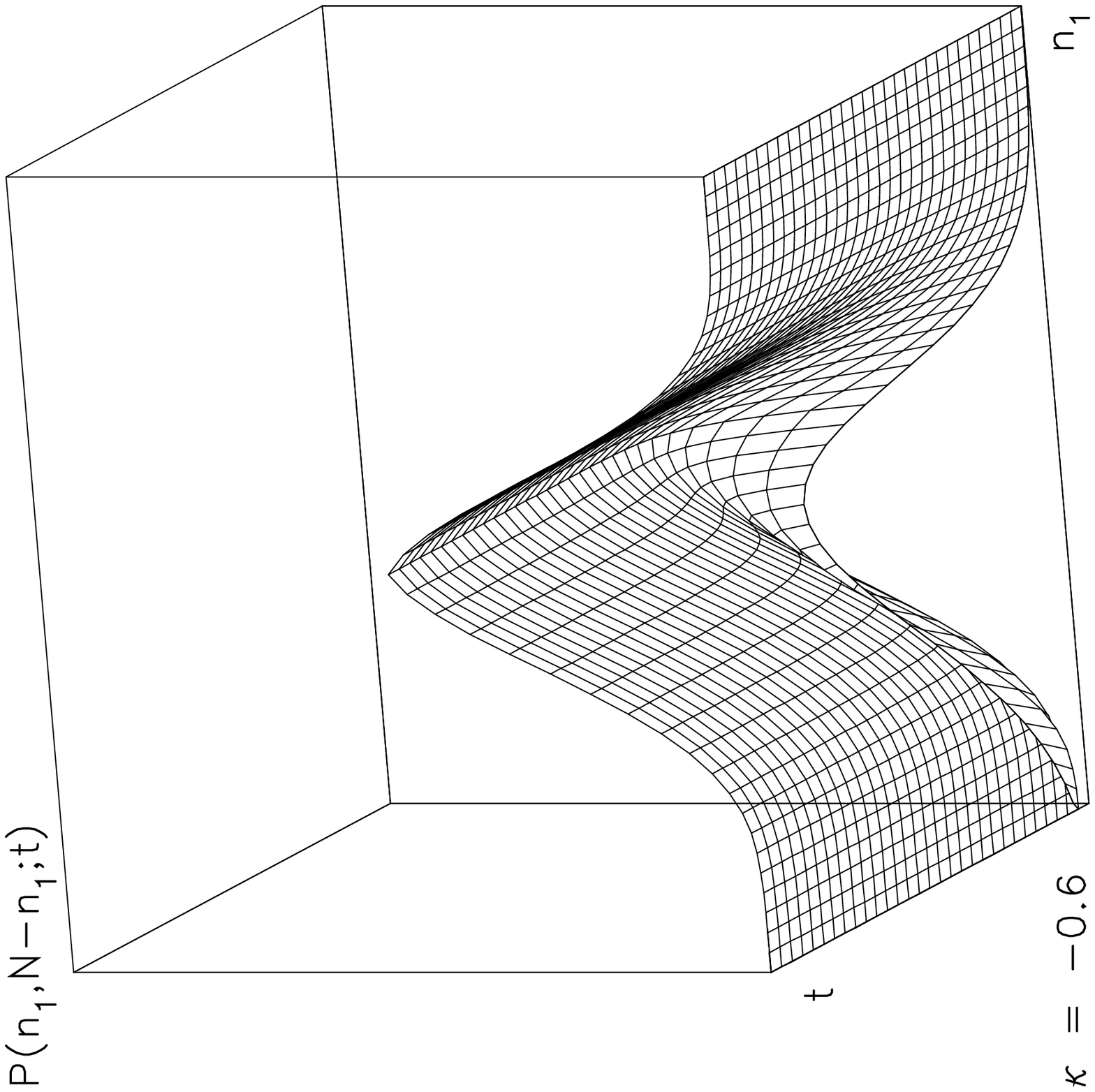}}}}
}\hfill
\parbox[b]{7cm}{
\epsfxsize=7cm 
\centerline{\rotate[r]{\hbox{\epsffile[28 28 570
556]{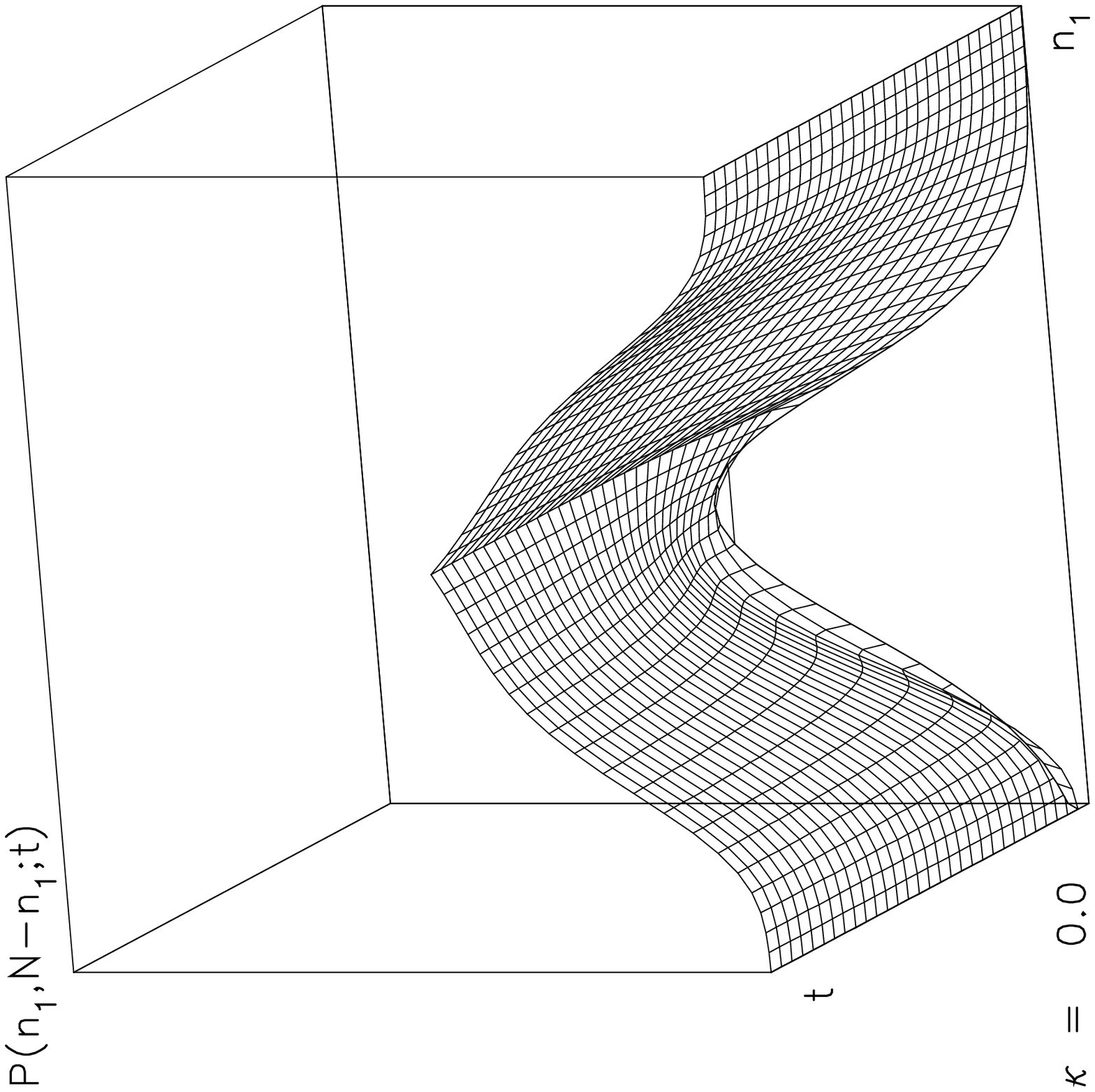}}}}
}                                              
\parbox[b]{7cm}{
\epsfxsize=7cm 
\centerline{\rotate[r]{\hbox{\epsffile[28 28 570
556]{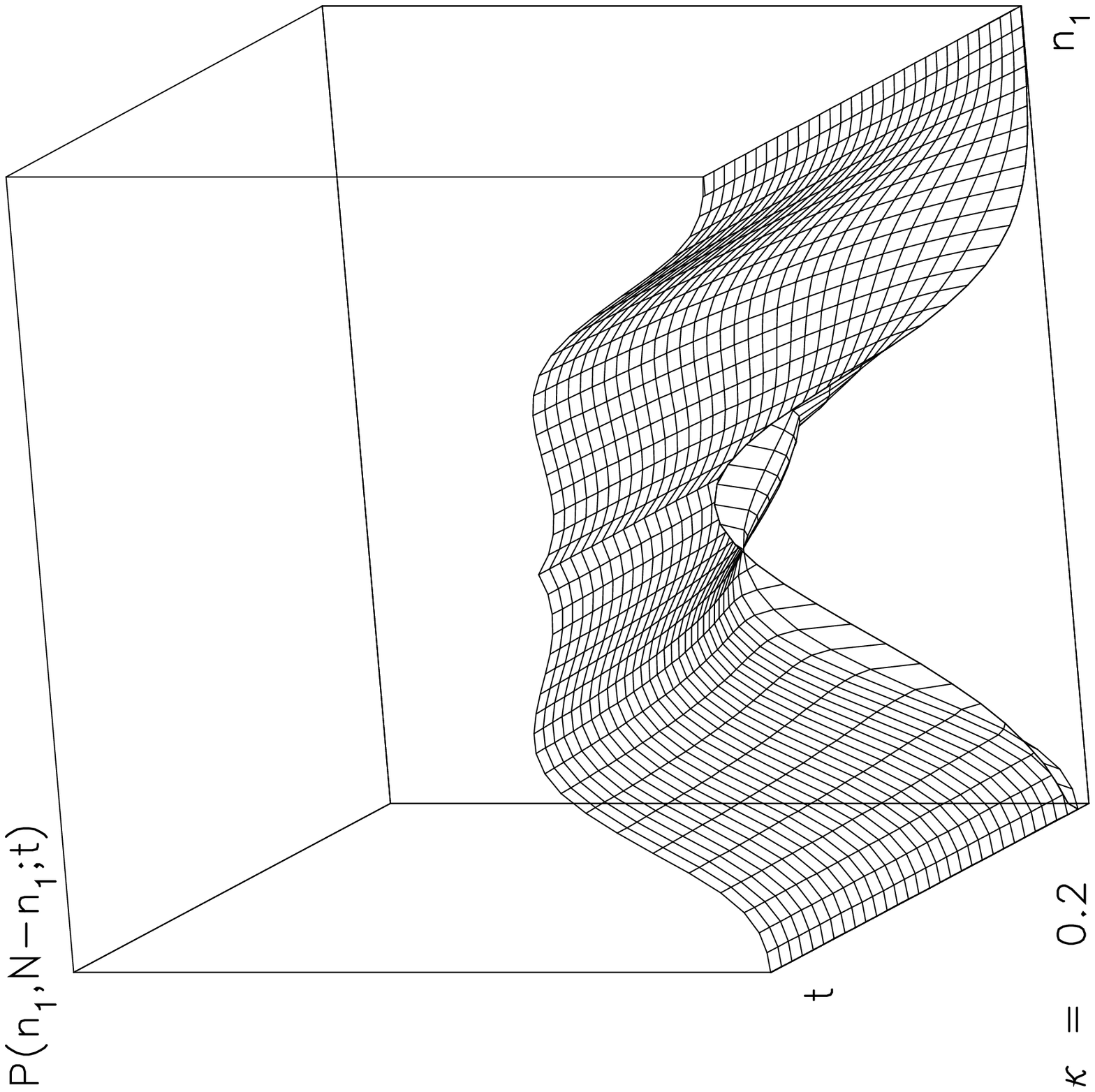}}}}
}\hfill
\parbox[b]{7cm}{
\epsfxsize=7cm 
\centerline{\rotate[r]{\hbox{\epsffile[28 28 570
556]{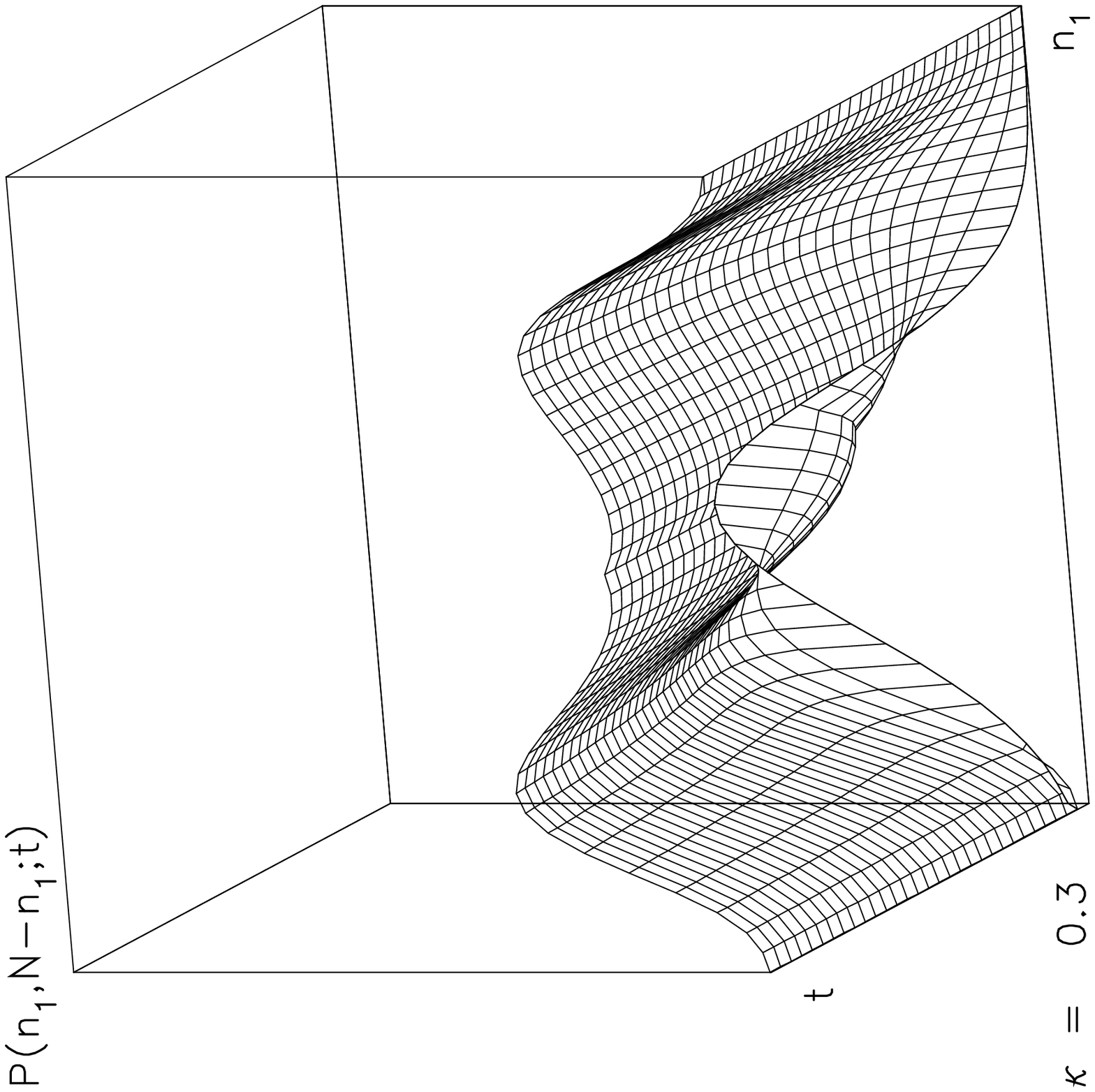}}}}
}
\caption[]{Probability distribution $P(\vec{n},t)
= P(n_1,N-n_1;t)$ of the socioconfiguration $\vec{n}$
for varying values of the control parameter $\kappa$ according to the
stochastic version of the game dynamical equations.\label{F1}}
\end{figure}

It is very interesting how the above mentioned symmetry breaking affects
the probability distribution $P(\vec{n},t) = P(n_1,n_2,t)
=P(n_1, N-n_1,t)$ of the related stochastic model
(cf. Fig. \ref{F1}\footnote{For illustrative reasons, a small number 
of individuals ($N=40$) and a broad initial probability distribution
have been chosen. In each picture, the box is twice as high as the maximal 
occuring value of the probability.}).
For $\kappa < 0$ the probability distribution is located around $n_1 
= N/2 = n_2$ and stays small so that the approximate mean value equations 
are applicable.
At the so-called {\it critical point} $\kappa =0$, 
a {\it phase transition} to a qualitative different
system behavior occurs and the probability distribution becomes very
broad. As a consequence, the game dynamical equations do not correctly describe
the temporal evolution of the mean strategy distribution 
anymore.
\par
For $\kappa > 0$, a bimodal and symmetrical probability distribution evolves.
That is, the likelihood that one of the two equivalent strategies will
win through is much larger than the likelihood to 
find approximately equal proportions
of both strategies. At the beginning, the initial state or maybe some
random fluctuation determines, which strategy has better chances to win.
However, in the long run both strategies have exactly the same chance.
It is clear, that in such cases the game dynamical equations fail to 
describe the mean system behavior (cf. Fig. \ref{F2}), 
which would correspond to the average 
temporal evolution of an ensemble of identical social systems. 
In cases of oscillatory or chaotic solutions
of the game dynamical equations the situation is even worse.

\section{Exact, Approximate, and Corrected Mean Values and Variances}

In the last section we have seen that the approximate mean value equations
\begin{equation}
 \frac{d\langle n_x \rangle_t}{dt}
 = M_x(\langle \vec{n} \rangle_t) 
\end{equation}
with the so-called {\it first jump moments}
\begin{equation}
 M_x(\vec{n}) = \sum_{\vec{n}^{\,\prime}}
 (n'_x - n_x) W(\vec{n} \rightarrow \vec{n}^{\,\prime})
\end{equation}
(cf. (\ref{not}))
are not sufficient. This calls for corrected mean value equations and
a criterium for the time period of their validity.
If the individual transition rates $w(x \rightarrow y;\vec{n})$ depend
on the socioconfiguration, the exact mean value can only be evaluated
via formula (\ref{av}). This requires the calculation of the probability 
distribution $P(\vec{n},t)$ and, therefore, the 
numerical solution of the respective master equation (\ref{master}).
Since the number of possible socioconfigurations is normally very large,
an extreme amount of computer time would be necessary for this. 
\par
Luckily, it is possible to derive from (\ref{not}) the
{\it corrected mean value equations}
\begin{equation}
 \frac{\partial \langle n_x \rangle_t}{\partial t}
 = M_x(\langle \vec{n} \rangle_t) + \frac{1}{2} \sum_y \sum_{y'}
 \sigma_{yy'}(t) \frac{\partial^2 M_x(\langle \vec{n} \rangle_t)} 
 {\partial \langle n_y \rangle_t \partial \langle n_{y'} \rangle_t} 
\label{corrmean}
\end{equation}
by means of a suitable Taylor approximation. This equation depends on the
{\it covariances}
\begin{equation}
 \sigma_{xy}(t) = \Big\langle \big( n_x - \langle n_x \rangle_t\big)
 \big(n_y - \langle n_y \rangle_t\big) \Big\rangle_t
 = \sum_{\vec{n}} \big(n_x - \langle n_x \rangle_t\big)
 \big(n_y - \langle n_y \rangle_t\big) P(\vec{n},t)
\end{equation}
which can be determined by means of the {\it covariance equations}
\begin{eqnarray}
 \frac{\partial \sigma_{xx'}(t)}{\partial t}
 &=& M_{xx'}(\langle \vec{n} \rangle_t) + \frac{1}{2}
 \sum_y \sum_{y'} \sigma_{yy'}(t) \frac{\partial^2 M_{xx'}(\langle \vec{n}
 \rangle_t)}{\partial \langle n_y \rangle_t \partial \langle
 n_{y'} \rangle_t} \nonumber \\
 &+& \sum_y \left[ \sigma_{xy}(t) \frac{\partial M_{x'}(\langle \vec{n}
 \rangle_t)}{\partial \langle n_y \rangle_t} + \sigma_{x'y}(t)
 \frac{\partial M_x(\langle\vec{n}\rangle_t)}{\partial \langle n_y \rangle_t}
 \right] \, . 
\label{corrcov}
\end{eqnarray}
The functions
\begin{equation}
 M_{xy}(\vec{n}) = \sum_{\vec{n}^{\,\prime}}
 (n'_x - n_x) (n'_y - n_y ) W(\vec{n} \rightarrow \vec{n}^{\,\prime})
\end{equation}
are called the {\it second jump moments}. 
\par
\begin{figure}[htbp]
\unitlength8mm
\begin{center}
\begin{picture}(16,10.6)(0,-0.8)
\put(0,9.8){\epsfig{height=16\unitlength, width=9.8\unitlength, angle=-90, 
      bbllx=50pt, bblly=50pt, bburx=554pt, bbury=770pt, 
      file=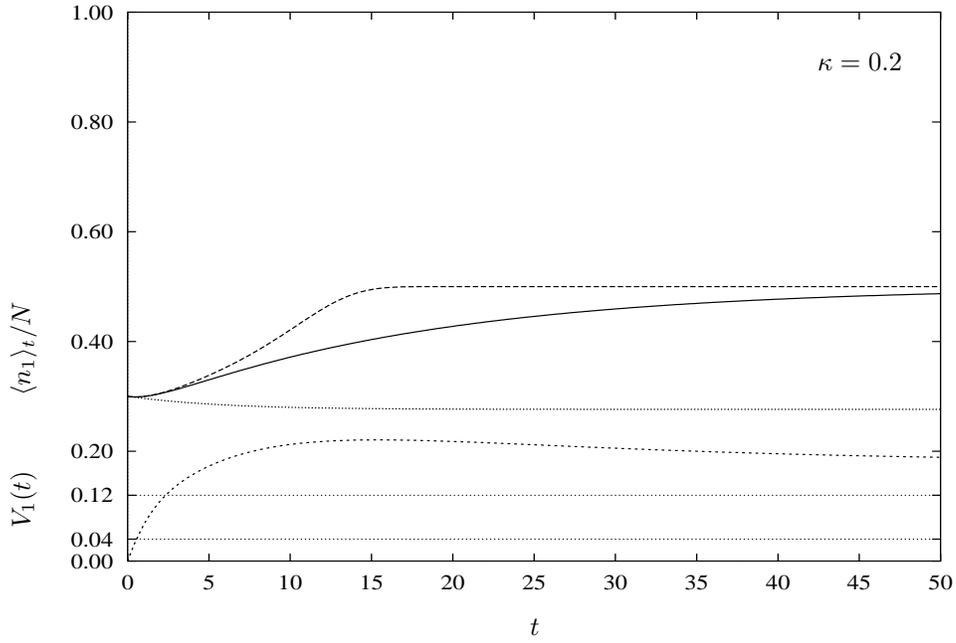}}
\put(8.7,-0.6){\makebox(0,0){\footnotesize $t$}}
\put(14.1,8.8){\makebox(0,0){\footnotesize $\kappa = 0.2$}}
\put(0.2,5.1){\makebox(0,0){\rotate[l]
{\hbox{\footnotesize $V_1(t)$\hspace*{1cm}$\langle n_1 
\rangle_t/N$\hspace*{3.5cm}}}}}
\end{picture}
\end{center}
\caption[]{The numerical solutions of the 
approximate mean value equations ($\cdots$) agree with those of the 
exact mean value equations (---) only for a short time interval. 
The corrected mean value equations (-- --) yield much better results,
although they also deviate from the exact curves when 
the relative variances (- - -) become too large. Nevertheless, they
describe the average long-term behavior properly.\label{F2}}
\end{figure}

Equations (\ref{corrmean}) and (\ref{corrcov}) 
build a closed system of equations, but
still no exact one, since this would depend on higher moments of the
form $\langle n_x n_y n_z \cdots \rangle_t$. Nevertheless, according to
Figure \ref{F2} the corrected mean value equations yield significantly better
results than the approximate ones. As a consequence, they are
valid for a much longer time period. Suitable 
{\it validity criteria} are the {\it relative variances}
\begin{equation}
 V_x(t) := \frac{\sigma_{xx}(t)}{(\langle n_x \rangle_t)^2} \, ,
\end{equation}
since these are a measure for the relative width of the
probability distribution $P(\vec{n},t)$. It can be shown that
the covariances and all higher moments are small, if only 
$V_x(t)$ is much smaller than 1 for every $x$.
Numerical investigations
indicate that the approximate mean value equations begin to separate from
the exact ones as soon as one of the relative variances
$V_x(t)$ becomes greater than 0.04.
The corrected mean value equations and covariances remain reliable as long
as $V_x(t)$ is smaller than 0.12 for all $x$ (cf. Fig. \ref{F2}).
\par
A more detailled discussion of the above matter
is presented elsewhere\footnote{D. Helbing,
``A Stochastic Behavioral Model and a `Microscopic' Foundation of
Evolutionary Game Theory'', in {\it Theory and Decision} 40, 1996, 
pp. 149--179.}.

\section{Diverse Generalizations}

The above discussed behavioral model can be generalized in different
respects. 

\begin{figure}[htbp]
\parbox[b]{7cm}{                                                   
\epsfxsize=7cm 
\centerline{\rotate[r]{\hbox{\epsffile[28 28 570                              
556]{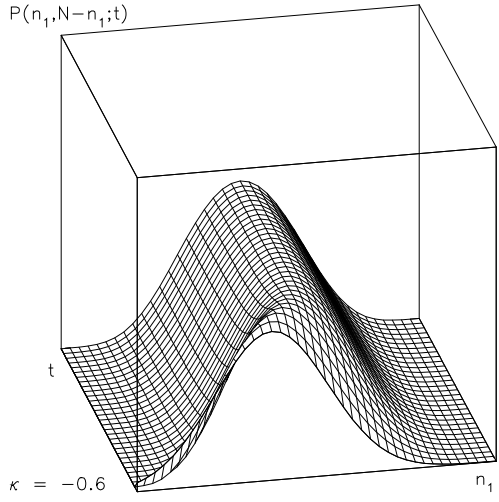}}}}
}\hfill
\parbox[b]{7cm}{
\epsfxsize=7cm 
\centerline{\rotate[r]{\hbox{\epsffile[28 28 570
556]{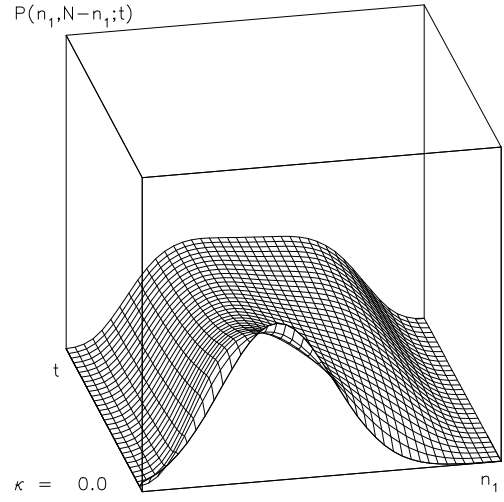}}}}
}                                              
\parbox[b]{7cm}{
\epsfxsize=7cm 
\centerline{\rotate[r]{\hbox{\epsffile[28 28 570
556]{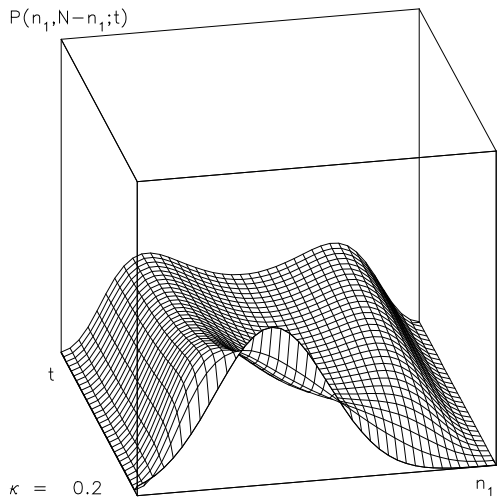}}}}
}\hfill
\parbox[b]{7cm}{
\epsfxsize=7cm 
\centerline{\rotate[r]{\hbox{\epsffile[28 28 570
556]{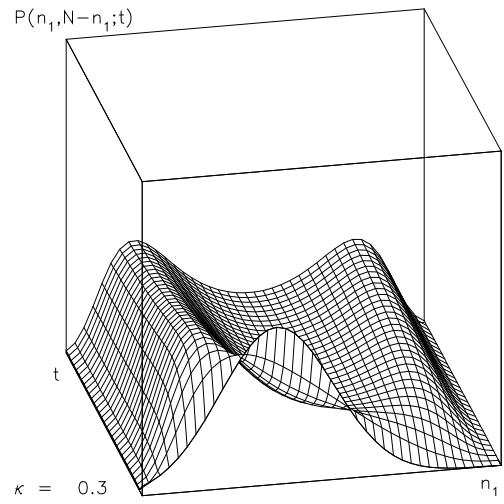}}}}
}
\caption[]{Probability distribution $P(\vec{n},t)
= P(n_1,N-n_1;t)$ of the socioconfiguration $\vec{n}$
according to the modified stochastic game dynamical equations.\label{F3}}
\end{figure}

\paragraph{Modified transition rates:}
The strange cusp at $n_1 = N/2$ in Figure \ref{F1}, which
comes from the discontinuous derivative of $w_2(x\rightarrow y)$ at
$E_x = E_y$, can be avoided by the modified imitation rates
\begin{equation}
 w_2(y \rightarrow x) = \frac{\nu}{N} \, \frac{\exp(E_x - E_y)}
 {D_{xy}} \qquad \mbox{with} \qquad D_{xy} = D_{yx} = 2 \, .
\end{equation}
This ansatz agrees with relation (\ref{prop2}) in linear approximation
for $C = \nu/(2N)$ and $\lambda = 1/2$, but it always yields non-negative
imitation rates. Similar to (\ref{prop}) it
guarantees two essential things: 1. The imitation rate grows with an
increasing gain $(E_x - E_y)$ 
of success. 2. If the alternative strategy $x$ is inferior, the imitation rate
is very small (but, due to uncertainty, not negligible).
The results of the corresponding stochastic behavioral model
are presented in Figure \ref{F3}. They show the usual flatness of the
probability distribution $P(n_1,N-n_1,t)$ at the critical point
$\kappa = 0$, where again a phase transition occurs.

\paragraph{Dynamics with expectations:} The decisions of individuals are
often influenced by their {\it expectations} $\langle E_x \rangle_{t'}^*$ about the success
of a strategy $x$ at future times $t' > t$. These will base on
some kind of extrapolation of past experiences with the
success of $x$. If expected payoffs at future times $t'$ 
are weighted exponentially with their distance $(t'-t)$ from 
the present time $t$, one would set\footnote{N. S. Glance and B. A. Huberman,
``Dynamics with Expectations'', in {\it Physics Letters A} 165, 1992,
pp. 432--440.}
\begin{equation}
 \langle E_x \rangle_t = \frac{1}{T} \int\limits_t^\infty dt' \,
 \langle E_x \rangle_{t'}^* \, \exp \left( \frac{t' - t}{T} \right) \, .
\end{equation}

\paragraph{Other kinds of pair interactions:} Apart from
imitative behavior, individuals also sometimes show an {\it avoidance
behavior}
\begin{equation}
 x + x \rightarrow y + x \, ,
\end{equation}
especially if they dislike their interaction partner (so-called `snob
effect'). This can be taken into account by an additonal contribution to
the individual interaction rates:
\begin{equation}
  w(y \rightarrow x;\vec{n}) = w_1(y \rightarrow x) + 
 w_2(y \rightarrow x) \, n_x + w_3(y \rightarrow x) \, n_y \, .
\end{equation}
$w_3$ denotes the {\it avoidance rate}.

\paragraph{Several subpopulations:}
Sometimes one has to distinguish different {\it subpopulations} $a$,
i.e. different kinds of individuals. This is necessary, if not all
individuals have the same set $S$ of strategies.\footnote{J. Hofbauer and
  K. Sigmund, {\it The Theory of Evolution and Dynamical Systems}.
Cambridge: Cambridge University Press 1988.} 
A similar thing holds, if
the considered social system consists of competing groups, where only 
individuals of the same group behave cooperatively. The generalized
behavioral equations are\footnote{D. Helbing, {\it Quantitative Sociodynamics.
Stochastic Methods and Models of Social Interaction Processes.} Dordrecht:
Kluwer Academic 1995.}
\begin{equation}
 \frac{dp_x^a(t)}{dt} = \sum_y \big[ p_y^a(t) w^a(y\rightarrow x;
 \langle \vec{n} \rangle_t ) 
 - p_x^a(t) w^a(x\rightarrow y;\langle \vec{n} \rangle_t ) \big] 
\end{equation}
with individual interaction rates of the form
\begin{equation}
  w^a(y \rightarrow x;\vec{n}) = w_1^a(y \rightarrow x) + 
 \sum_b \big[ w_2^{ab}(y \rightarrow x) \, n_x^b 
 + w_3^{ab}(y \rightarrow x) \, n_y^b \big] \, .
\end{equation}

\paragraph{Inclusion of memory effects:}
If the strategy distribution at past times $t' < t$ influences 
present decisions in a non-Markovian way, the approximate mean value
equations have the form
\begin{equation}
 \frac{dp_x^a(t)}{dt} = \sum_y \int\limits_{-\infty}^t dt' \,
 \big[ p_y^a(t') w_{t-t'}^a(y\rightarrow x;
 \langle \vec{n} \rangle_{t'} ) 
 - p_x^a(t') w_{t-t'}^a(x\rightarrow y;\langle \vec{n} \rangle_{t'} ) 
 \big] \, .
\end{equation}
For example, in cases of an exponentially decaying memory one would have
\begin{equation}
 w_{t-t'}^a(x\rightarrow y;\langle \vec{n} \rangle_{t'} )
 = w^a(x\rightarrow y;\langle \vec{n} \rangle_{t'} )
 \, \frac{1}{\tau} \, \exp \left( \frac{t - t'}{\tau} \right) \, .
\end{equation}

\section{Summary and Conclusions}

We have found a microscopic foundation of the game dynamical equations,
basing on a certain kind of imitative behavior. Moreover, a stochastic version
of evolutionary game theory has been formulated. It allowed to understand
the self-organization of social conventions as a phase transition which is
related with symmetry breaking. Moreover, we have seen that 
the game dynamical equations correspond to approximate mean value 
equations. Normally, they agree with the mean value 
equations of stochastic game theory
for a certain time period only, which can be determined by calculating
the relative variances. For an improved description of the average system behavior
we have derived corrected mean value equations which require the
solution of additional covariance equations.
\par
The interpretation of the game dynamical equations
follows by reformulating these in terms of a {\it social 
force model,}\footnote{D. Helbing, {\it Stochastische Methoden, nichtlineare
Dynamik und quantitative Modelle sozialer Prozesse,} 2nd edition. 
Aachen: Shaker 1996.} assuming a continuous strategy set:
\begin{equation}
 \frac{dx_\alpha(t)}{dt}  = f_1(x_\alpha) + \sum_{\beta (\ne \alpha)} f_2
 (x_\alpha,x_\beta) + \mbox{\it fluctuations} \, .
\label{force}
\end{equation}
The force term
\begin{equation}
 f_1(x_\alpha) = \int dx \, (x - x_\alpha) \, w_1(x_\alpha 
 \rightarrow x) 
\end{equation}
delineates spontaneous strategy changes by individual $\alpha$, whereas
\begin{eqnarray}
 f_2(x_\alpha,x_\beta) &=& 
 (x_\beta - x_\alpha) \, w_2(x_\alpha \rightarrow x_\beta) \nonumber \\
 &+& \int dx \, ( x - x_\alpha) \, w_3(x_\alpha \rightarrow x) \, 
 \delta (x_\alpha - x_\beta)
\end{eqnarray}
is the {\it interaction force} which originates from individual $\beta$ 
and influences individual $\alpha$. Here, $\delta(x - y)$ denotes
{\it Dirac's delta function} (which yields a contribution for $x=y$ only).
According to (\ref{force}), the game dynamical equations describe the
{\it most probable strategy changes} rather than the average 
(representative) evolution of a social system. 
Therefore, they neglect the effects of fluctuations on the
system behavior. 
\par
A more detailled discussion of the results presented in this paper is
available elsewhere$^{12,13}$.
\clearpage
\pagebreak
\pagebreak
\end{document}